\documentclass[preprint,showpacs,amsmath,amssymb,prb]{revtex4}
\usepackage{graphicx}

\begin{document}

\title{Thermal Conductivity of Nanotubes Revisited:
Effects of Chirality, Isotope Impurity, Tube Length, and Temperature}
\author{Gang Zhang}
\email{gangzh@stanford.edu}
\affiliation{Department of Physics, National University of Singapore, Singapore 117542, Republic of Singapore}
\affiliation{Department of Chemical Engineering, Stanford University, CA 94305-4060 }
\author{Baowen Li}
\email{phylibw@nus.edu.sg}
\affiliation{Department of Physics, National University of Singapore, Singapore 117542, Republic of Singapore}
\affiliation{NUS Graduate School for Integrative Sciences and Engineering, 117597, Republic of Singapore}

\date{J. Chem. Phys. {\bf 123}, 114714 (2005)}

\begin{abstract}
We study the dependence of thermal conductivity of single walled
nanotubes (SWNT)  on chirality, isotope impurity, tube length and
temperature by nonequilibrium molecular dynamics method with
accurate potentials. It is found that, contrary to electronic
conductivity, the thermal conductivity is insensitive to the
chirality. The isotope impurity, however, can reduce the thermal
conductivity up to $60\%$ and change the temperature dependence
behavior. We also found that the tube length dependence of thermal
conductivity is different for nanotubes of different radius at
different temperatures.
\end{abstract}

\maketitle

Carbon nanotube is one of exciting nano-scale materials discovered
in  the last decade. It reveals many excellent mechanical, thermal
and electronic properties\cite{textbook}.  Depends on its
chirality\cite{elec1,elec2}, the nanotube can be either metallic
or semiconducting. For example, for zigzag (9,0) and (10,0) tubes,
their radius are almost the same, but (9,0) tube behaves metallic
and (10,0) tube semiconducting. At room temperature, the
electronic resistivity is about $10^{-4}-10^{-3}\Omega \mbox{cm}$
for the metallic nanotubes, while the resistivity is about
$10\Omega \mbox{cm}$ for semiconducting tubes \cite{resistivity}.
The $10\%$ difference in radius induces the change of electronic
conductivity  in four orders of magnitude. One may ask, whether
thermal conductivity is also very sensitive to the chirality like
its electronic counterpart?

On the other hand,  the isotope impurity reduces thermal
conductivity of most materials, such as germanium and
diamond\cite{ge,diamond1}. It is surprising that $1\%$ $^{13}$C in
diamond leads to a reduction of thermal conductivity up to
$30\%$\cite{diamond2}. Is there same effect in carbon nanotubes?

Moreover, in electronic conductance, SWNT reveals many 1D
characters\cite{textbook}.  However in thermal conduction, it is
still not clear whether the conduction behavior is like that one
of 1D lattice or a quasi 1D (1D lattice with transverse motions)
or that one in a 2D lattice.

These questions and many other relevant properties of nanotubes
are very important and should be understood before the nanotubes
are put into any practical application. Indeed, recent years have
witnessed increasing interesting in thermal conductivity of
nanotubes\cite{tc1,tc2,tc3,tc4,tc5,tc6,tc7a,tc7b,tc8,tc9,tc10,tc11},
the questions raised here are still open.

In this paper,  we study the effects of the chirality, isotope
impurity, tube length and temperature on SWNTs' thermal
conductivity  by using the non-equilibrium molecular dynamics (MD)
method with bond order potential. This approach is valid as it
shows that at finite temperature, phonon has a dominating
contribution to thermal conduction than electron
does\cite{heat1,zigzag}. We should pointed out that the thermal
conductivity calculated in this paper is exclusively from lattice
vibration. Of course, for the metallic nanotubes, the electrons
may give some but limited contributions\cite{heat1,zigzag}, but
this is not the main concern in our paper.

The Hamiltonian of the carbon SWNT is:
\begin{equation}
H=\sum_i\left( \frac{p_i^2}{2m_i}+V_i\right) ,\quad V_i=\frac
12\sum_{j,j\neq i}V_{ij}
\end{equation}
where $V_{ij}=f_c(r_{ij})[V_R(r_{ij})+b_{ij}V_A(r_{ij})]$ is the Tersoff
empirical bond order potential. $V_R(r_{ij})$, and $V_A(r_{ij})$ are the
repulsive and attractive parts of the potential, and $f_c(r)$ depending on
the distance between atoms. $b_{ij}$ are the so-called bond parameters
depending on the bounding environment around atoms $i$ and $j$, they
implicitly contain many-body information. Tersoff potential has been used to
study thermal properties of carbon nanotubes successfully\cite{tc6,poten1}.
For detailed information please refer to Ref\cite{Tersoff}.

In order to establish a temperature gradient, the two end layers
of nanotube are put into contact with two Nos\'e-Hoover heat
bathes \cite{Heatbath} with temperature $T_{L}$ and $T_{R}$ for
the left end and the right end, respectively. Free boundary
condition is used. All results given in this paper are obtained by
averaging about $10^6 \sim 10^7$ femtosecond (fs) after a
sufficient long transient time (usually $10^6 \sim 10^7$ fs) when
a non-equilibrium stationary state is set up and the heat flux,
$J$, becomes a constant. The thermal conductivity, $\kappa $, is
calculated from the Fourier law,
\begin{equation}
J=-\kappa \nabla T,
\label{FOURIER}
\end{equation}
where $J$ is defined as the energy transported along the tube in
unit time through unit cross section area, and $\nabla T=dT/dx$ is
the temperature gradient. In this paper, we chose $d=1.44{\AA}$ as
the tube thickness, thus the cross section is $2\pi r d$, where
$r$ is radius of the tube. In the following we shall discuss the
chirality dependence, isotope impurity and tube length effect.

\begin{figure}
\includegraphics[width=\columnwidth]{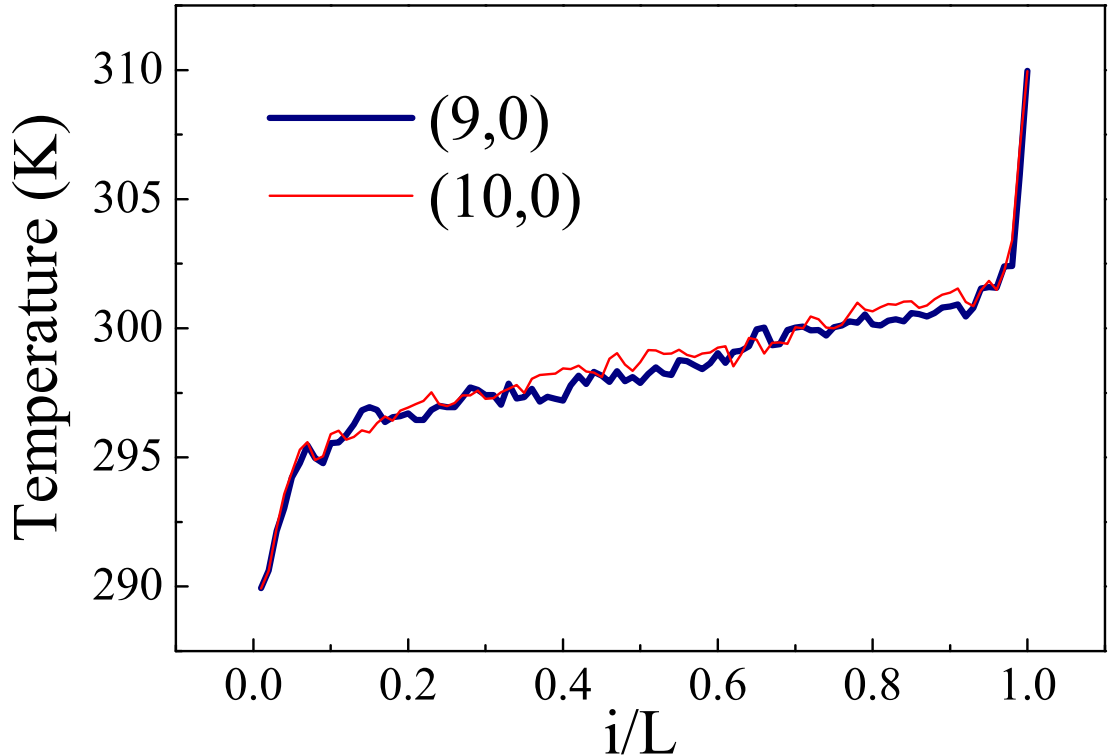} \vspace{-1.cm}
\caption{The temperature profiles of (9,0) and (10,0) SWNT at
$300K$.  The temperature is obtained by averaging over $5\times
10^6$ fs after dropping the $10^6 \sim 10^7$ fs transient time.
The tube length, $L$, is $108 {\AA }$.} \label{fig:zigzag}
\end{figure}

\textit{Chirality Dependence.} The thermal conductivity of zigzag
and armchair SWNTs of same length but  with different radius are
calculated. Fig. \ref
{fig:zigzag} shows the temperature profiles of (9,0) and (10,0) nanotubes at $%
300K$. These two temperature profiles are very close to each
other, thus the temperature gradient, $dT/dx$,  is almost the
same. The thermal conductivity, tube radius and relative thermal
conductance are listed in Table 1. The difference in thermal
conductivity comes mainly from radius difference. If we use
thermal conductance (defined as thermal conductivity times cross
section area), the relative value (to (9,0) tube) for (9,0),
(10,0) and (5,5) SWNTs with the same length is $1:0.97:1.05$.

It is clear that, unlike its electronic counterpart, the thermal
conductivity/conductance of SWNTs does not depend on the chirality
and/or atomic geometry sensitively both at low temperature and
room temperature. Our MD results are consistent with that one from
Landauer transmission theory\cite{zigzag}.  The electron DOS of
SWNTs depends on chirality. There is an energy gap at Fermi level
in $(10,0)$ tube, while no such a gap in $(9,0)$ tube. However,
the phonon DOS's in different tubes do not show any significant
difference\cite{phonon}.

\begin{table}
\caption{Thermal conductivity for different tubes.}
\label{tb-1}%
\begin{ruledtabular}
\begin{tabular}{ccccc}
\\
 \mbox{Tube}& $(9,0)$ &$(10,0)$ & $(5,5)$\\
$\kappa$  $(\mbox{W/mK})$ &$880$  & $770$ & $960$ \\
$\mbox{Radius, r}$ $(\AA )$ &$3.57$  & $3.97$ & $3.43$ \\
$\mbox{Relative}$ $\mbox{thermal}$ $\mbox{conductance}$ &$1.0$  & $0.97$ & $1.05$ \\
\end{tabular}
\end{ruledtabular}
\end{table}

\textit{Isotope impurity effect.}
Isotope impurity affects many physical properties of materials, such
as thermal, elastic, and vibrational properties\cite{isotope1}.
There are three isotopes of carbon element, $^{12}$C, $^{13}$C, and $^{14}$C.
They have the same electronic structure, but different masses. Here we
study the effect of $^{14}$C impurity on thermal conductivity
of SWNTs. In our calculations, $^{14}$C atoms are randomly distributed in a $%
^{12}$C SWNT. To suppress the possible fluctuations arising from random
distribution, an average over $10$ realizations is performed for each
conductivity calculation. We also do the calculation for tube with $^{13}$C impurity and similar effect is found.

\begin{figure}
\includegraphics[width=\columnwidth]{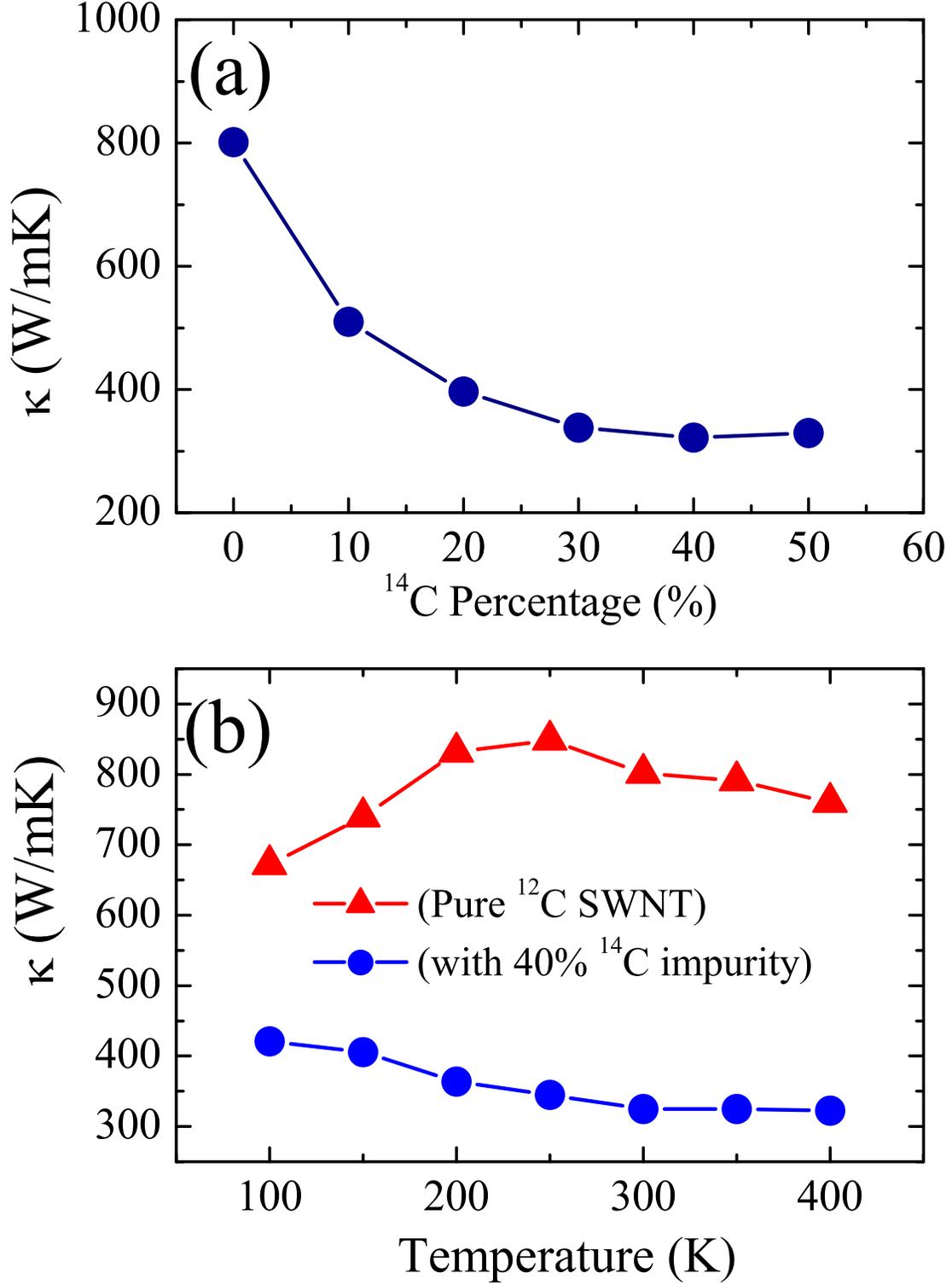} \vspace{-1.cm}
\caption{(a) Thermal conductivity, $\kappa$, versus $^{14}$C impurity
percentage for a (5,5) SWNT at $300K$. (b) Thermal conductivity, $\kappa$,
versus temperature for a (5,5) pure $^{12}$C nanotube ( solid $\triangle$) and a (5,5) SWNT with $40\%$ $%
^{14}$C impurity ($\bullet$). The curves are drawn to guide the eye. }
\label{fig:isotop}
\end{figure}

Fig. \ref{fig:isotop}(a) shows the dependence of thermal
conductivity on the impurity percentage. We select armchair
$(5,5)$ tube with a fixed length of about $60 {\AA}$. The thermal
conductivity  decreases as the percentage of $^{14}$C impurity
increases. With $40\% - 50\%$ $^{14}$C, the thermal conductivity
is reduced to about $40\%$ of that one in a pure $^{12}C$ SWNT.
This is similar to the isotope effect on thermal conductivity of
diamond\cite{diamond2}.

The thermal conductivity decreases more quickly at low percentage
range than  at high range. From this curve, we can estimate
roughly that the thermal conductivity decreases about $20\%$ with
only $5\%$ $^{14}$C isotope impurity. This decrease is not as
rapid as that one in a diamond that $1\%$ isotope impurity can
reduce  thermal conductivity as much as $30\%$\cite{diamond2}.

This result tells us that one can modulate the thermal
conductivity of carbon nanotubes by adding $^{14}$C or other
isotope impurity as it alters only the thermal conductivity  and
has no effect on the electronic properties.

 The thermal conductivity, $\kappa $, for (5,5) pure $^{12}$C nanotube and that one with $40\%$ $%
^{14}$C impurity at different temperatures with same tube length
are shown  in Fig. \ref{fig:isotop}(b). The difference for these
two cases are obvious. The isotope impurity changes completely the
temperature dependence behavior of thermal conductivity. For a
pure tube (solid $\triangle$), there is a maximum at about $T_M
\approx 250$K. Below this temperature, $\kappa $ increases when
$T$ is increased. Above $T_M$, $\kappa$ decreases with increasing
$T$. However, in the case with isotope impurity, there is no
maximum in the curve. The thermal conductivity monotonically
decreases as the temperature increases.

These phenomena can be understood from the phonon scattering
mechanism.   Increasing temperature has two effects on thermal
conductivity. On the one hand, the increase of temperature will
excite more high frequency phonons that enhance thermal
conductivity. We call this effect ``positive" effect. On the other
hand, the increase of temperature will also increase phonon-phonon
scattering that in turn will increase the thermal resistance, thus
suppress the energy transfer. We call it `negative" effect. The
thermal conductivity is determined by these two effects that
compete with each other.

For a pure SWNT, at low temperature regime, the phonon  density is
small, the ``positive" effect dominates, thus  the thermal
conductivity increases with increasing temperature.  However, at
high temperature regime, as more and more (high frequency) phonons
are exited, the ``negative" effect dominates, which results in the
decrease of thermal conductivity as temperature is increased. The
results in Fig \ref{fig:isotop}(b) is consistent the results from
Savas et al.\cite{tc6}.

However in the case with impurity, the scattering mechanism
changes. In this case, most high frequency phonons are localized
due to the impurity. The main contribution to heat conduction
comes from the low energy phonon that has long wavelength. The
``positive" effect is largely suppressed, and in the whole
temperature regime, the ``negative" effect dominates that leads to
a decrease of thermal conductivity as the temperature is
increased, as is seen in Fig \ref{fig:isotop}(b).

\textit{Tube length effect.} Recent years' study on heat
conduction in low dimensional lattices shown that for a one
dimensional lattice without on-site potential, thus momentum is
conserved, the thermal conductivity $\kappa$ diverges with system
size (length)\cite{FPU}, $L$, as $\kappa\sim L^{\beta}$, with
$\beta=2/5$. If the transverse motions are allowed, like in the
quasi 1D case\cite{0.33}, then $\beta=1/3$. Moreover, it has been
found that this anomalous conduction is connected with the
anomalous diffusion\cite{LW03}. For more detailed discussion on
anomalous heat conduction and anomalous diffusion in different
models such as lattice models, billiard gas channels, and
nanotubes, see Ref\cite{LWWZ05}. However, in a 2D system, it is
still not known at all (both analytically and numerically) that
what shall be the divergence form: power law form or logarithmic
form or anything else.

As for the SWNT, Maruyama\cite{tc7a,tc7b} has studied the
$\kappa(L)$ for tubes of different radius, and found
that\cite{tc7b} the value of $\beta$ decreases from $0.27$ for
(5,5) tube, to 0.15 for (8,8) and 0.11 for (10,10).

Here we investigate this problem from different aspects,  namely,
we study the change of $\beta$ in different temperatures. For
comparison, we also study a 1D carbon lattice model with the same
interatomic potential. The carbon-carbon atom distance is also
$1.44{\AA }$.

\begin{figure}
\includegraphics[width=\columnwidth]{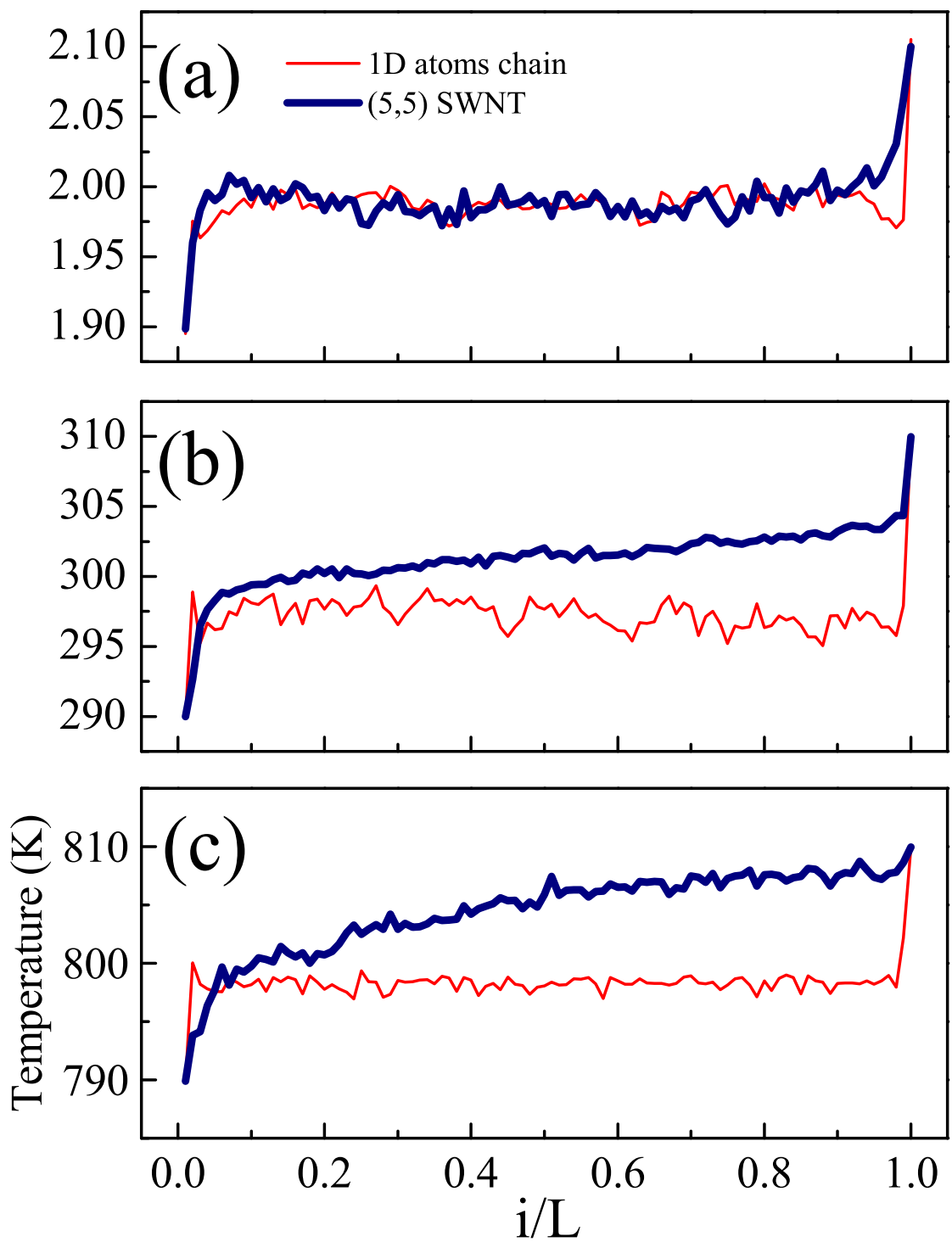}\vspace{-.7cm}
\caption{The temperature profile of (5,5) SWNT and 1D lattice at (a)
2K, (b) 300K, and (c) 800K. The temperature is obtained by averaging over $5\times 10^6$ fs
after dropping the $10^6 \sim 10^7$ fs transient time.}
\label{fig:T-profile}
\end{figure}

The temperature profiles for a carbon SWNT and a 1D carbon lattice are shown in Fig \ref
{fig:T-profile} for different temperatures. In
this figure, both SWNT and 1D lattice has $100$ layers of atoms. Fig.\ref
{fig:T-profile}(a) demonstrates clearly that at low temperature as low as $%
2$K, there is no temperature gradient in both the SWNT and 1D lattice. This
resembles the 1D lattice model with a harmonic interaction potential\cite
{Lebowitz}. This can be understood from the Taylor expansion of the Tersoff
potential by keeping up to the second order term. Because at low
temperature, the vibrations of atoms are very small, the potential can be
approximated by a harmonic one. And in SWNT, the vibration displacement in
transverse direction is much smaller than the one along the tube axis and
can be negligible. This result means that energy transports ballistically in
SWNTs at low temperature.

\begin{figure}
\includegraphics[width=\columnwidth]{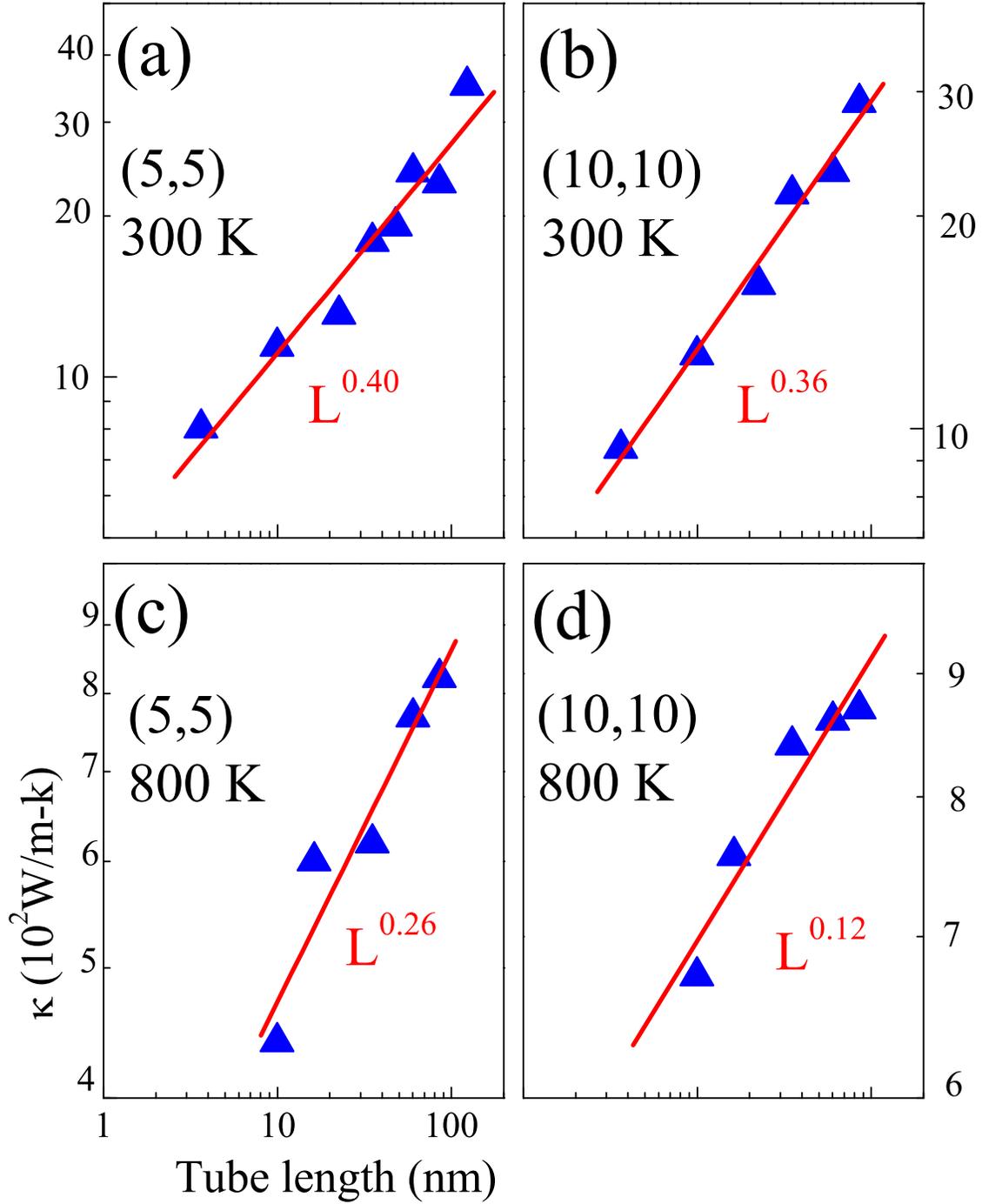}\vspace{-.7cm}
\caption{The thermal conductivity, $\kappa$, versus tube length,
$L$,  in log-log scale for (5,5) and (10,10) tubes at  300K and
800K. In all cases, $\kappa \sim L^{\beta}$ with $\beta$ changes
from case to case.
 The solid line, whose slope is the value of $\beta$, is the best fit one. }
\label{fig:kappa}
\end{figure}

However at high temperature, the situation changes. As is shown in
Fig. \ref{fig:T-profile}(b) for $300$K, and Fig.
\ref{fig:T-profile}(c) for  $800$K, there is still no temperature
gradient in the 1D lattice, but in the SWNT, temperature gradient
is set up. In 1D lattice with Tersoff potential, the increase of
temperature does not change the harmonic character; while in the
SWNT at room temperature and higher temperature, transverse
vibration  increases as temperature increases. The temperature
gradient is established due to the interaction of the transverse
modes and longitudinal modes.

The thermal conductivity, $\kappa $, versus the tube length, $L$,
is shown in log-log scale in Fig.\ref{fig:kappa}(a-d) for (5,5)
and (10,10) SWNT at 300K and 800K, respectively. Obviously, the
value of $\beta $ depends on temperature as well as tube radius.
For a SWNT, $\beta $ decreases as temperature increases (cf. (a)
and (c), (b) and (d)); and at the same temperature, $\beta $
decreases as the tube radius increases (cf. (a) and (b), (c) and
(d)). This can be qualitatively explained by the modes coupling
theory\cite{0.33}. At high temperature, the transverse vibrations
are much larger than that at low temperature, thus interaction
between the transverse modes and longitudinal modes becomes
stronger, which leads to a smaller value of $\beta$. For (5,5)
tube at T=300 K, because the small tube radius, the transverse
modes can help set up the temperature gradient but the coupling
with the longitudinal mode is still very weak, and the
longitudinal modes dominate the heat conduction, this is why in
this case the thermal conduction behaviour is very close to that
one in the 1D Fermi-Pasta-Ulam type lattices\cite{FPU}, namely,
the thermal conductivity, $\kappa$, diverges with $L$, as
$L^{0.4}$.

It is worth pointing out that the absolute values of thermal
conductivities given in Fig. 4 for (5,5) and (10,10) tubes are
different from that ones given in Refs. \cite{tc7a,tc7b}. The
reasons are that: (a) the wall thickness used in
Refs.\cite{tc7a,tc7b} was 3.4$\AA$, while in our case we use
1.44$\AA$, this leads to our results are at least about 2.5 times
larger than those ones in Refs\cite{tc7a,tc7b}; (b) the
thermalstate we used are different from that one in
Refs.\cite{tc7a,tc7b}; (c) boundary conditions are different, we
use free boundary condition while Refs\cite{tc7a,tc7b} use fixed
boundary condition. If we use conductance (get rid of the effect
of tube thickness and tube radius) rather than the conductivity,
we believe that our results are consistent with that one from
others\cite{tc7a,tc7b,tc11}

In summary, we have studied the effects of chirality,  isotope
impurity, tube length, and tube radius on thermal conductivity  in
SWNTs. Our  results show that the thermal conductivity is
insensitive to the chirality. However, the introduction of isotope
impurity suppresses thermal conductivity of carbon nanotubes up to
$60\%$ and change the temperature dependence behavior. Moreover,
at low temperature the heat energy transfers ballistically like
that one in a 1D harmonic lattice, while at high temperature,
thermal conductivity diverges with tube length,  $L$, as
$L^{\beta}$. The value of $\beta$ depends on temperature and tube
radius. This unique structural characteristic makes SWNT an ideal
candidate for testing heat conduction theory, in particular, the
mode-coupling theory\cite{0.33} in low-dimensional systems.

This project is supported in part by a Faculty Research Grant of
National University  of Singapore and the Temasek Young
Investigator Award of DSTA Singapore under Project Agreement
POD0410553 (BL) and Singapore Millennium Foundation (GZ).

\end{document}